\documentclass[10pt]{article}
\begin{document}
\title{Note on Varying Speed of Light Cosmologies}
\author{George F\ R\ Ellis. \\
Mathematics Department and Applied Mathematics,\\
University of Cape Town, South Africa.}
 \maketitle

\begin{abstract}
\noindent The various requirements on a consistent varying speed
of light (`VSL') theory are surveyed, giving a short check-list of
 issues that should be satisfactorily handled by such theories.
\end{abstract}

\section{Varying Speed of Light (`VSL') Cosmologies}
Many papers have been published on varying speed of light (`varying
\emph{c}') theories, with claims \emph{inter alia} that they can
solve some of the problems that inflationary cosmologies seek to
solve. However many of those papers are problematic, and the
increasing literature in the area does not fully take cognisance of
some foundational issues for any VSL theory.

This note will not comment on any specific such papers in detail,
although it will use a few as examples of what is discussed. Rather
it will aim to provide the reader with a toolkit enabling a critique
of any VSL paper, checking if it meets some basic criteria that can
be claimed to be fundamental for an adequate discussion of VSL
proposals. A key point will be distinguishing theories where it is
the speed of light $v_{photon}$ - the speed at which photons travel
- that is supposed to vary, and those where it is a universal
causally limiting speed $v_{lim}$ that is supposed to vary. While
these are the same speed $c$ in standard relativity theory, they can
and indeed probably will differ in any genuine VSL theory.

One might ask why it is necessary to place an article containing
material that seems rather elementary in a research journal. The
reason is that VSL papers that ignore these fundamentals are
continually being put on the archive and even being published in
reputable journals. This paper will hopefully help future work in
the area avoid common pitfalls in formulating such theories.

\section{Fundamentals: the speed of light and measurement}
First, note that just because a quantity is called `the speed of
light' and labeled `c' does not necessarily mean it is the speed of
light in physical terms. What determines that some quantity is in
fact the physical speed of light $v_{photon}$? In standard
relativity theory \cite{hawell73} this is related on the one had to
the metric tensor, which determines time and spatial measurements as
well as the geometry of null geodesics, and on the other hand to
Maxwell's equations, which determine the paths of light rays in
space time. These are linked because the characteristics of wavelike
solutions of Maxwell's equations are null geodesics, as determined
by the metric tensor. Thus in standard General Relativity (`GR'), it
is a fundamental interpretational principle that light rays - the
paths of photons and other massless particles in spacetime - are
solutions of the equation
\begin{equation}
ds^2 \equiv g_{\mu\nu}(x^k)dx^\mu dx^\nu = 0, \label{eq:g1}
\end{equation}
which therefore determines the speed of light $v_{photon}$.

Second, meaningful variations in the constants of nature refer
only to dimensionless constants, because only dimensionless
quantities have an invariant meaning when units of measurement are
changed. The quantity `c' is not dimensionless; consequently, one
can change the apparent speed of light to any desired value by
changing the coordinates or the units used. This does not mean the
physical speed of light has changed. Indeed one can always set the
speed of light to unity by appropriate choice of units: simply
rescale either spatial distances or time units, as one prefers,
and one can set any physically measured speed of light to unity.
Thus for example if the metric of Minkowski spacetime is initially
given as
\begin{equation}
ds^2 = - dt^2 + dr^2/c^2,\,\,dr^2 \equiv dx^2 + dy^2 + dz^2
\label{eq:g2}
\end{equation}
then `c' is indeed the speed of light $v_{photon}$, as (\ref{eq:g1})
implies
\begin{equation}
ds^2 = 0 \Rightarrow (v_{photon})^2 = dr^2/dt^2 = c^2. \label{eq:c1}
\end{equation}
Now one can rescale: $X = x/c$, $Y = y/c$, $Z = z/c$, to get
\begin{equation}
ds^2 = - dt^2 + dR^2, \,\, dR^2 \equiv  dX^2 + dY^2 + dZ^2,
\label{eq:g3}
\end{equation}
and the speed of light has been set to unity:
\begin{equation}
ds^2 = 0 \Rightarrow (v_{photon})^2 = dR^2/dt^2 = 1, \label{eq:c2}
\end{equation}
 without any loss of generality of the physics involved. By similar
 transformations we can make the speed of light anything we want;
this is not a puzzle, as of course this is just the coordinate speed
of light.

The key issue relating this to physics is how one measures spatial
distances and times, for it is only when we have distance and time
units set up that we can start to measure the speed of light. One
can claim \cite{ellwil00} that on macroscopic scales - from meters
to solar system scale - the only practical way of determining
distance is via radar; other astronomical distance scales are
derived from this one.\footnote{Parallax distance measurements for
example rely on knowing the physical size of the base used to
determine the parallax; and that has to be determined by some method
such as radar.} One assumes a good clock can be constructed, and
then uses the timing of reflected electromagnetic radiation to
determine the distance. But then the (physical) speed of light of
necessity has to be unity, precisely because all electromagnetic
radiation travels at the speed of light, and distances are being
determined by use of such radiation. This is reflected in the
natural units used for such distance measurements: light seconds and
light years. When such units are used, the speed of light is unity
by definition - not by definition of how fast light moves, but by
the definition used for spatial distances.\footnote{Indeed Synge has
emphasized that time is \emph{the} fundamental measurement:
\emph{all} classical physical quantities can be expressed in terms
of units of time (\cite{syn60}, see particularly Appendix B.)} From
a practical point of view, the speed of light has been fixed by the
BIPM (Bureau International des Poids et Mesures), and we derive the
meter from it. Then equation (\ref{eq:g1}) does not \emph{determine}
the speed of light - it \emph{defines} it.

\noindent\textbf{Implication 1}: \emph{In order to be viable, any
VSL theory involving a variable speed of photon travel must of
necessity be based on some other method of measuring spatial
distances than radar. So the question for any specific proposed VSL
theory is, What viable alternative proposal for distance measurement
is made?}\footnote{The majority of VSL papers do not discuss this
issue.}

\section{The speed of light and the metric}
In terms of the metric tensor components, in standard relativity
theory we are allowed to make basis transformations that are
\emph{not} Lorentz transformations; and they can make the metric
tensor components anything we want! This is essentially what is
being done in cosmology when the Robertson-Walker metric tensor
\begin{eqnarray}
ds^2 &=& - c_0^2 dt^2 + a^2(t) d\sigma^2,\,\, d\sigma^2 \equiv dr^2
+ f^2(r) (d\theta^2 + \sin^2\theta d\phi^2) \label{eq:m1}
\end{eqnarray}
is assumed to hold initially, and then postulated to change to
\begin{eqnarray}
ds^2 &=& - c_1^2 dt^2 + a^2(t) d\sigma^2 \label{eq:m2}
\end{eqnarray}
at some critical time $t_*$ in the early universe, where $c_i$ are
constants.\footnote{A `phase change' is supposed to take place. This
is essentially what is proposed in \cite{mof02}, see eqns.
(18)-(20).} Then (\ref{eq:m1}) implies on radial null geodesics
\begin{eqnarray}
ds^2 = 0 \Rightarrow v_{photon} = (1/c_0)a(t) dr/dt, \label{eq:c3}
\end{eqnarray}
while (\ref{eq:m2}) implies on radial null geodesics
\begin{eqnarray}
ds^2 = 0 \Rightarrow v_{photon} = (1/c_1) a(t) dr/dt, \label{eq:c4}
\end{eqnarray}
giving an apparent change in the speed of light by a factor
$c_0/c_1$. However in fact the units of measurement have been
changed (leading to a consequent change in the metric tensor
components), rather than the physical speed of light altering.
Indeed one can transform (\ref{eq:m2}) to (\ref{eq:m1}) by the
change of coordinates $t \rightarrow (c_0/c_1) t$. So according to
the principle of general covariance, they are just the same
spacetime in different coordinates (this feature being confused by
using the same label `t' for what are in fact two different time
coordinates used at earlier and later times). The two metrics do not
represent different physical speeds of light.

Is there some preferred time coordinate that can break this
degeneracy in the coordinate speed of light? Yes indeed; proper time
$\tau$ is by definition the time measured along a (timelike) world
line by a perfect clock. In both standard special relativity (`SR')
and GR, it is give by the line integral
\begin{eqnarray} \tau = \int d\tau, \, \, d\tau \equiv \sqrt{-ds^2}
\, = \, \sqrt{-g_{\mu\nu}(x^k)dx^\mu dx^\nu}. \label{eq:m3}
\end{eqnarray}
Applying this to the fundamental world lines with tangent vector
$u^a = \delta^a_0/c_0$ in (\ref{eq:m1}) reveals that in that metric
proper time along these world lines is $\tau = c_0 t$. Similarly
(\ref{eq:m2}) reveals the proper time $\tau$ along these world lines
is $\tau = c_1 t$. The quantity (\ref{eq:m3}) is an invariant, and
will be the same whatever coordinate system is used. While we can
use any coordinates, some are more convenient than others in that
they more directly represent the physics of what is going on; we get
these preferred coordinates on choosing the time coordinate $t$ as
proper time $\tau$ along the fundamental world lines at all times.
Then $c_0 = 1$ and $c_1=1$, (\ref{eq:m1}) is identical to
(\ref{eq:m2}), and there is no jump in the apparent speed of light.
If there is a spacetime where the metric tensor components change at
some time $t_*$ from the form (\ref{eq:m1}) to (\ref{eq:m2}) with
$c_o \neq c_1$, nothing has changed physically; there has simply
been a rescaling in the time coordinate used. This has not affected
the physical speed of light.

Similarly spatial distances are determined by the integral of the
line element $\sqrt{ds^2}$ along spacelike curves. This prescription
together with (\ref{eq:m3}) gives a relation between time and
spatial distance measurements compatible with the radar definition
of distance mentioned in the previous section, and so is a way of
codifying the feature that the speed of light is invariant in
standard GR.

\noindent \textbf{Implication 2}: \emph{Any varying speed of light
theory based on changes in the metric tensor components must explain
how it differs from GR as regards how time and space measurements
are related to the metric tensor. What replaces (\ref{eq:m3}) in the
proposed theory?}

\section{The speed of light, causality, and the Lorentz group} The
speed of light plays a key role in standard physics because it is
the limiting speed $v_{lim}$ for local relative motion, as indicated
by the standard relativistic laws for velocity transformations
derived from the Lorentz group. This is indicated by the way it is a
universal speed, invariant under velocity addition:
\begin{equation}\label{add}
 \forall v:  \, v + v_{lim} \rightarrow v_{lim}.
\end{equation}
It is consequently a speed that cannot be exceeded by any physical
object or information-carrying signal, and this is related for
example to the increase of inertial mass with relative motion,
diverging as this speed is approached. Thus $v_{lim}$ determines
local causality. This basically occurs because physics is Lorentz
invariant, and this limiting speed is uniquely characterized by the
fact that it is Lorentz invariant; indeed this is what underlay the
initial derivation of the Lorentz group.

The link to the metric tensor is that Lorentz transformations are
precisely those transformations that leave the metric tensor
components invariant. A key role is played by the relation between
the contravariant and covariant metric tensor components:
\begin{eqnarray}
g_{ab}  g^{bc} = \delta_a^c \label{eq:gg}
\end{eqnarray}
which relates the speed of light in these two sets of components:
using Synge's convention \cite{syn60} that time measurements will be
fundamental, at any point the metric components can be written as
\begin{eqnarray} g_{ab}  = diag\{-1,\, 1/c^2,\,
1/c^2,\,1/c^2\}\, \Rightarrow \,\,g^{ab}  = diag\{-1,
\,c^2,\,c^2,\,c^2\}. \label{eq:m4}
\end{eqnarray}
Lorentz transformations preserve both these canonical metric tensor
forms. The limiting causal speed $v_{lim}$ is just the `speed of
light' constant $c$ that occurs in both these metric forms and is
left invariant by the Lorentz group. It consequently occurs in the
Lorentz factor
\begin{equation}\label{gamma}
\gamma(v) = 1/ \sqrt{1-v^2/c^2}
\end{equation}
characterizing the kinematic effects of length contraction and time
dilation as well as the dynamic effect of increasing inertial mass;
indeed the standard time dilation formula follows immediately from
the conjunction of (\ref{eq:m3}) with (\ref{eq:m4}) (see
\cite{ellwil00}), and the dynamical limits on speed follow from the
relativistic increase in mass: $m = m_0 \gamma(v)$. But this
limiting speed can always be set to unity, globally in SR as
explained above (see the transformation from (\ref{eq:g2}) to
(\ref{eq:g3}) above), and locally at any point in GR (in essence,
because of the principle of equivalence together with the special
relativity result). Natural units for spatial distances and time,
associated with normalized coordinates and associated orthonormal
bases of vectors, indeed give this normalized form (\ref{eq:g3}) at
any point:
\begin{eqnarray} g_{ab}  =
diag\{-1,\, 1,\, 1,\,1\}, \,\,g^{ab}  = diag\{-1, \,1,\,1,\,1\},
\label{eq:m5}
\end{eqnarray}
showing explicitly that the limiting speed $v_{lim}$ (which is the
quantity `c' in (\ref{eq:m4}) and (\ref{gamma})) is unity; as is
required for consistency, $\gamma(v) = 1/\sqrt{1-v^2}$.

In the cosmological context, we can if we wish change the time
variable $t$ in the metric (\ref{eq:m1}) to $\eta = \int dt/a(t)$ to
obtain the conformally flat metric form
\begin{eqnarray}
ds^2 &=& a^2(t) \{- d\eta^2 + d\sigma^2\}\label{eq:mc1}.
\end{eqnarray}
By (\ref{eq:g1}), radial light rays plotted in terms of the
coordinates $(\eta, r)$ will always be at $\pm 45^o$; thus the
limiting speed $v_{lim}$ is again found to be unity. These are the
coordinates used in the standard Penrose conformal diagrams of
cosmology, used to make clear causal relations in these spacetimes
\cite{ellwil00}. Changing time or space coordinates does not affect
these causal relations, which can always be represented by such
diagrams with $c=1$.\footnote{Hence Figure 2 of \cite{albmag99}
shows some light cones as flatter than $45^0$ because unusual
coordinates have been chosen. Choice of conformal coordinates would
bring this diagram to standard form with all light rays at $\pm
45^o$, and hence no variation of the speed of light. The same issue
arises with Figure 2 of \cite{mag03a}.}

In SR and GR, the equations for all fundamental physical fields are
written in Lorentz invariant form. The consequent use of the metric
tensor\footnote{For example in raising and lowering indices.} in the
action or in resultant field equations such as the Klein-Gordon
equation implicitly introduces the speed of light $c$, ensuring that
causality works out correctly. The wavelike characteristics for
massless physical fields are then light rays, because the wave
operator
\begin{eqnarray}
\nabla^2 f  &=& g^{ab} f_{;ab} \label{eq:wave}
\end{eqnarray}
relates these characteristics to the metric tensor, and hence to the
value of $c$ occurring in that tensor. Discontinuities in these
fields and waves will travel at the speed of light as specified by
the metric tensor components (\ref{eq:m4}), and their causal
properties will be represented correctly by the Penrose conformal
diagrams mentioned above.

\noindent\textbf{Implication 3}: \emph{Any varying speed of light
theory involving a change in the limiting speed will not be Lorentz
invariant; the way Lorentz invariance is broken must be made
explicit}.\footnote{Such a proposal is made in \cite{albmag99} for
example,and see \cite{mag03a} for discussion.} \emph{If it involves
an alteration of the Lorentz group, it must make clear how the
limiting dynamical speed is determined}.

\section{The speed of light and Maxwell's equations}
Of particular concern in relation to the speed of light is the
physics that in fact determines the speed of light. Any adequate
varying speed of light theory cannot just postulate \emph{ad hoc}
changes to the speed of light $v_{photon}$, perhaps assuming it can
undergo a `phase transition' for example, without some physical
reason for such a change; doing so is not proposing a coherent
physical theory. The physics supposed to underlie the variation of
$v_{photon}$ should be made explicit.

At a minimum one can (in the spirit of present day physics) assume
$c$ is a dynamical variable and propose some effective field
equations to govern its variation. Any such equations must
eventually be derivable either from Maxwell's equations or from
whatever other equations will be taken as underlying electromagnetic
theory. One can perhaps in an investigative spirit leave that
problem for the moment and just assume some effective equations to
govern its time and space variation. But that can only be a
temporary expedient: eventually if it is indeed the speed of light
$v_{photon}$ we are talking about, it must be related to the
equations of electromagnetism.\footnote{See \cite{luchod90} for a
clear discussion disentangling the link between SR and
electromagnetism.}

In standard relativity theory, the metric tensor is used to raise
and lower indices on the Maxwell field tensor $F_{ab}$, and hence
relates the two sets of Maxwell's equations:
\begin{eqnarray}
 F_{[ab;c]} =0,
\,\,F^{ab}_{\,\,\,\, ;b} = 0,\,\,  F^{ab} \equiv g^{ac} g^{bd}
F_{bd},\label{eq:f}
\end{eqnarray}
thus by (\ref{eq:m4}) inserting $c$ in the derivative terms in these
equations.\footnote{It may also occur in the relation between the
tensor $F_{ab}$ and the electric and magnetic fields $E$, $B$ as
well as the charge $e$ and current $j$. However these occurrences of
$c$ are dependent on the units used for $E$, $B$, $e$ and $j$, and
can be absorbed in those quantities without loss of generality.} The
divergence relation $F^{ab}_{\,\,\,\, ;ab} = 0$ then leads to the
standard wave equation for electromagnetism, with $c$, deriving from
the metric tensor form (\ref{eq:m4}), being the physical speed of
light $v_{photon}$, and so corresponding to a massless photon. As
pointed out above, $c$ can be normalised to unity in the metric form
without loss of generality, and then the actual speed of light is a
constant rather than a variable quantity.

One can avoid this conclusion in various ways: for example utilising
a bimetric theory with one metric determining space and time
measurements and another being used in Maxwell's
equations,\footnote{For a viable implementation of this idea, see
\cite{basetal00}.} invoking a non-minimal coupling between
electromagnetism and gravity,\footnote{For a viable implementation
of this idea, see \cite{tes04}.} or proposing a theory of massive
photons such as given for example by the Proca action \cite{poe05}.
The first key point is that \emph{some} proposal needs to be made in
this regard: when one talks about the speed of light varying, this
only gains physical meaning when related to Maxwell's equations or
its proposed generalisations, for they are the equations that
determine the actual speed of light. The second key point is that
just because there is a universal speed $v_{lim}$ does not prove
there is a particle that moves at that speed. On the standard
theory, massless particles move at that speed and massive particles
don't; but by itself, that result does not imply that massless
particles exist. Their existence is a further assumption of the
standard theory, related to the wavelike equations satisfied by the
Maxwell electromagnetic field. One can consider theories with
massive photons.

\noindent \textbf{Implication 4}: \emph{Any varying speed of light
theory involving a change in the speed of photon travel must
eventually propose some other equations than standard Maxwell's
equations to govern electromagnetism, and show how this leads to a
varying physical speed of light associated with a wavelike solution
to these equations.}

\section{The speed of light and dynamical equations}
Any proposed variation of the speed of light has major consequences
for almost all physics, as it enters many physics equations in
various ways, particularly because of the Lorentz invariance built
into fundamental physics (see \cite{elluza05} for discussion). One
needs to take cognisance of the effects of this variation on the
rest of physics.

What is sometimes done is to just allow the constant $c$ to become a
variable in one or more physical equations, and then proclaim this
as a variable speed of light theory. Three points are relevant here.
First, if a combination of `constants' varies in a dynamical
equation, it might be any of them that is responsible. It is only
specifically the speed of light that varies if it is related to the
physical speed of light as discussed in the previous sections. For
instance\footnote{See section V of \cite{albmag99} for an example.}
if one allows the quantity $c$ to become a variable in the equation
of state $p=w \rho c^2$, it is the combination $\hat{w} = w c^2$
that is the physically effective quantity that is made to vary.
Indeed one can absorb $c^2$ into the definition of the equation of
state variable $w$, as this just amounts to a choice of physical
units for $p$. Then the normalized equation of state is just $p =
\hat{w}\rho$, and what appeared to be a variable speed of light
theory can be regarded as just introduction of a time-dependent
equation of state. As another example \cite{barmag98}, if in
\begin{equation}\label{efe}
    G_{ab} = \kappa T_{ab} + \Lambda g_{ab},
\end{equation}
where
\begin{equation}\label{k}
\kappa \equiv \frac{8\pi G}{c^4}
\end{equation}
one allows the speed of light $c$ to become a variable, then
$\kappa$ becomes a variable and this changes the solutions to these
gravitational equations. Equivalently, one can propose such a
variation in the action
\begin{equation}\label{var}
    S = \int \sqrt{-g}\,dx^4\left(\frac{R + 2\Lambda}{2\kappa}+
    \textsl{L}_m\right)
\end{equation}
and work out the dynamic consequences. But the point then is that
the physical effects contemplated here occur only because $\kappa$
is now varying; and a varying $\kappa$ does not necessarily mean $c$
varies. Indeed if the supposed variation in $\kappa$ is not linked
to the other effects mentioned in the previous sections of this
note, then $c$ is not varying - because it is those effects that
identify the varying quantity to be the speed of light. Rather what
one has in fact done is propose a varying $G$ theory - and of course
there is a large literature on that topic. If it is to be
specifically $c$ that is varying, you must vary it in all places
where it occurs in the physical equations, which is many places; but
where it occurs in the equations depends on the units used, so it is
not obvious how to do this coherently and uniquely. For example in
the above equations we can define $\tilde{G} \equiv G/c^4$, which
amounts to a specific choice of units to be used for $G$; then
\begin{equation}\label{k1}
\kappa \equiv 8\pi \tilde{G}
\end{equation}
and $c$ does not appear in equations (\ref{efe}), (\ref{var}),
(\ref{k1}). Which form of the equations to use is a free choice; the
result of allowing $c$ to vary is quite different depending on the
form chosen.\footnote{A variant is allowing $c$ to vary in the
quantity $\rho_\lambda \equiv \Lambda c^2/8\pi G$, see (12)-(15) in
\cite{albmag99}; this is equivalent to letting either $\Lambda$ or
$G$ vary, as $c$ can be absorbed into the units used to measure
$\Lambda$.}

Secondly, when a quantity is changed from a constant to a
function, obtaining the correct dynamic equations requires doing
the variation with that quantity treated as a variable \emph{ab
initio}. One gets the wrong result by first doing the variation
and then changing the constant to a variable. For example, setting
$c^4 = \psi(x)$ in (\ref{k}) and adding a term $L_\psi$ into the
action (\ref{var}) results in a set of gravitational equations
that are not the same as those obtained by first performing the
variation and then setting $c^4 = \psi(x)$.\footnote{This is done
for example in \cite{barmag98}.} This is because the term
involving $R$ in (\ref{var}) is multiplied by $1/\kappa$, which is
then no longer constant. The first method is the one that gives
the correct result. Simply setting $c$ to a variable in the
dynamic equations after the variation has been performed leads to
problems with consistency of the resulting equations, see
\cite{basetal00,elluza05} for discussion.

Finally any such proposed change of $c$ in various physical
equations needs some motivation in physical terms: what fundamental
physical cause has changed the quantity $c$? This has been commented
on in the previous section. We just note here that any variable
which varies is either a combination of basic fields, or is itself a
field in the sense that it has a kinetic term. If one lets a
quantity vary in spacetime without giving it a kinetic term, this
violates what we know about basic physics.

\noindent \textbf{Implication 5}: \emph{Any varying speed of light
theory must be done consistently in terms of its effects on the
whole set of physical equations. You can't just allow $c$ to vary in
some equations and not in others; and the proposed variation in $c$
in these equations requires some viable physical explanation in
order to complete the set of causal relations.}

\section{An integrated whole}
The overall message is that you can't just alter the speed of light
in one or two equations and leave the rest of physics unchanged. It
plays a central role in modern physics \cite{elluza05}, particularly
because it is the invariant limiting speed of the Lorentz group and
so is built into any variables that transform under that group, but
also because electromagnetism is central to many physical effects;
in particular, light is central to measurement. On the standard
view, these various roles are tightly integrated together in a
coherent package in which the speed of light does not vary. Any
viable VSL theory has to propose a similarly integrated viable
alternative to the whole package of physical equations and
consequent effects (kinematical and dynamical) dependent on $c$. You
can't just tinker with a few bits of the whole.\\

\noindent \textbf{Acknowledgement}: I thank R Durrer and J-P Uzan
for helpful comments that have substantially improved this note.

\end{document}